\documentclass[twoside,leqno,twocolumn]{article}
\usepackage{ltexpprt}

\usepackage{graphicx}
\usepackage{amssymb}
\usepackage{amsmath}
\usepackage{algorithmicx}
\usepackage{algorithm}
\usepackage{algpseudocode}
\usepackage{pgfplotstable}
\usepackage{longtable}
\usepackage{multirow}
\usepackage{caption}
\usepackage{float, epsfig, epstopdf, balance, algorithm}
\usepackage[caption=false]{subfig}
\usepackage[hyphens]{url}
\usepackage[inline]{trackchanges}
\addeditor{JECA}
\addeditor{IVAN}
\addeditor{DJOLA}
\addeditor{ZOKI}

\newcommand*\samethanks[1][\value{footnote}]{\footnotemark[#1]}

\begin{document}

\title{Deep Attention Model for Triage of Emergency Department Patients}


\author{Djordje Gligorijevic\thanks{Authors contributed equally.} \thanks{Center for Data Analytics and Biomedical Informatics, Temple University, Philadelphia, PA, USA.} 
\and Jelena Stojanovic\samethanks[1] \samethanks[2] 
\and Wayne Satz\thanks{Lewis Katz School of Medicine, Temple University, PA, USA} \\
\and Ivan Stojkovic\samethanks[2] 
\and Kraftin Schreyer\samethanks[3] 
\and Daniel Del Portal\samethanks[3] 
\and Zoran Obradovic\samethanks[2]}
\date{}

\maketitle


\fancyfoot[R]{\footnotesize{\textbf{Copyright \textcopyright\ 2018 by SIAM\\
Unauthorized reproduction of this article is prohibited}}}





\begin{abstract} \small\baselineskip=9pt 
Optimization of patient throughput and wait time in emergency departments (ED) is an important task for hospital systems. For that reason, Emergency Severity Index (ESI) system for patient triage was introduced to help guide manual estimation of acuity levels, which is used by nurses to rank the patients and organize hospital resources. 
However, despite improvements that it brought to managing medical resources, such triage system greatly depends on nurse's subjective judgment and is thus prone to human errors. 
Here, we 
propose a novel deep model based on the word attention mechanism designed for predicting 
a number of resources an ED patient would need.
Our approach incorporates routinely available continuous and nominal (structured) data with medical text (unstructured) data, including patient's chief complaint, past medical history, medication list, and nurse assessment collected for     338,500 ED visits over three years in a large urban hospital. Using both structured and unstructured data, the proposed approach achieves the AUC of $\sim 88\%$ for the task of identifying resource intensive patients (binary classification), and the accuracy of $\sim 44\%$ for predicting exact category of number of resources (multi-class classification task), giving an estimated lift over nurses' performance by 16\% in accuracy. 
Furthermore, the attention mechanism of the proposed model provides interpretability by assigning attention scores for nurses' notes which is crucial for decision making and implementation of such approaches in the real systems working on human health.
\end{abstract}


\section{Introduction}
Due to patient overcrowding and increased acuity of waiting patients, the Emergency Departments (EDs) have been under ever-increasing pressure to improve utilization of their resources. For that reason, EDs of many hospitals in the US have been investing into improvements to their services by optimizing staff and resource requirements, patient wait time, and treatment outcomes \cite{gilboy1999,stojanovic2017}. The hospital management teams are highly motivated in improving the system efficiency, such that medical help can reach both the highest severity of illness and intensity of service groups, while also providing a separate queue for the least resource-intensive patients in an expedited manner\footnote{\url{https://www.acep.org/Clinical---Practice-Management/Utilization-Review-FAQ/}, accessed January 2018}.

An important aspect of the patient resource allocation in ED systems begin with the triage processes.
When patients arrive at the emergency department, they are processed at the triage area by a nurse who listens to their complaint and assesses acuity (i.e., urgency), completes entry of discrete history items plus writes a note summarizing the findings (e.g., patient's medical history and symptoms), and lastly measures basic body functions such as heart rate, blood pressure, or temperature. Then, once all relevant information is collected, it is used to assign each patient to a triage category. The practice of assessing patients' acuity levels and predicting the amount of required resources for treatment is called the {\it triage process}. Rating systems for acuity triage have traditionally been based solely on the illness severity of a patient, determined through the nurses assessment of vital signs, subjective and objective information, past medical history, allergies, and medication \cite{gilboy1999,gilboy2012}. Then, a nurse would assign the acuity level by making a judgment regarding severity of illness and intensity of service to help determine the acuity queue each patient is assigned to. The acuity assignment impacts the waiting time for each patient to transition to the next phase of care, typically the treatment area. However, this process was inherently flawed due to high variance and subjectivity of the practitioners, and variation in ability to correctly predict needed resources\cite{kosowsky2001}. 

The Emergency Severity Index \cite{gilboy2011} estimation of acuity has become the standard triage method over the past 15 years, and is the most dominant implemented system in the emergency departments in the United States at the moment \cite{mchugh2012}. The index has five levels of acuity \cite{wuerz2000}, where levels 1 and 2 are ranked as highly urgent and patients with such acuity are given the highest priority. On the other hand, the ESI levels of acuity 4 and 5 are considered non-urgent, and are often given lower priority or more commonly are placed in a separate queue. 
The ESI system was created with a goal to aid in patient triage and to help separate more complex (or resource-intensive) patients from those with simpler problems, and thus improve patient throughput and disposition decision\cite{gilboy2011}. 
The ESI triage improved over traditional systems by introducing 
nurses' estimation of the number of resources that a less acute patient would need during his/her ED stay. Resources may include the number of lab tests (e.g., blood or urine), ECG, X-ray, CT, MRI, or therapeutic interventions like fluids hydration or medications\footnote{Examples of Resources for ESI Levels 3-5. October 2014. Agency for Healthcare Research and Quality, Rockville, MD. \url{http://www.ahrq.gov/professionals/systems/hospital/esi/esitab4-2.html}, accessed October 2017}. Even consultation with a specialist is considered as a resource of emergency department\footnote{\url{http://www.ahrq.gov/professionals/systems/hospital/esi/esi4.html}, accessed October 2017}.

\begin{figure}[h!]
	\centering
	\captionsetup{justification=centering}
	\includegraphics[width=0.7\linewidth]{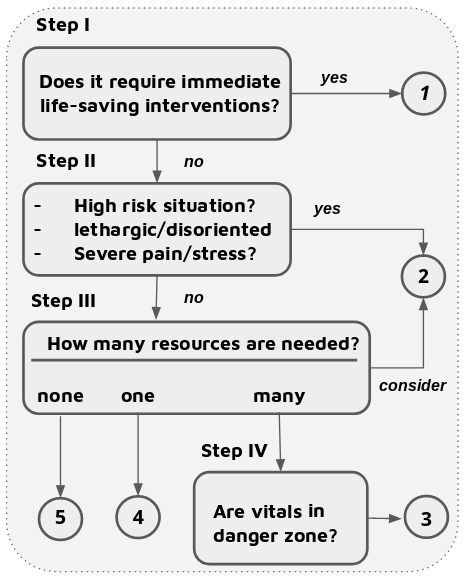}
	\caption{Emergency Severity Index (ESI): A Triage Tool for Emergency Department} 
	\label{fig:ESI}
\end{figure}

Note that resource prediction is only used for less acute patients. At decision steps I and II on the ESI algorithm (Figure~\ref{fig:ESI}), the nurse decides which patients meet criteria for ESI levels 1 and 2 based only on patient severity of illness. However, at decision step III, the nurse assigns ESI levels 3 to 5 by assessing both acuity and predicted resource needs. Thus, the triage nurse only considers resources when the answers to decision Step I and II are "no."
Correct estimation of resource consumption has significant consequences, as it was shown that patients who require two or more resources have higher rates of hospital admission and mortality, as well as longer lengths of stay in the ED \cite{eitel2003,tanabe2004}. As such, the ESI system was created with an assumption that triage nurses would be able to accurately estimate the number of resources for an ED stay using the flow algorithm shown in Figure~\ref{fig:ESI} and discriminate high acuity patients from low acuity ones. 

\begin{figure}[h!]
\centering
\subfloat[First acuity level patients distribution]{
\includegraphics[width=.35\textwidth]{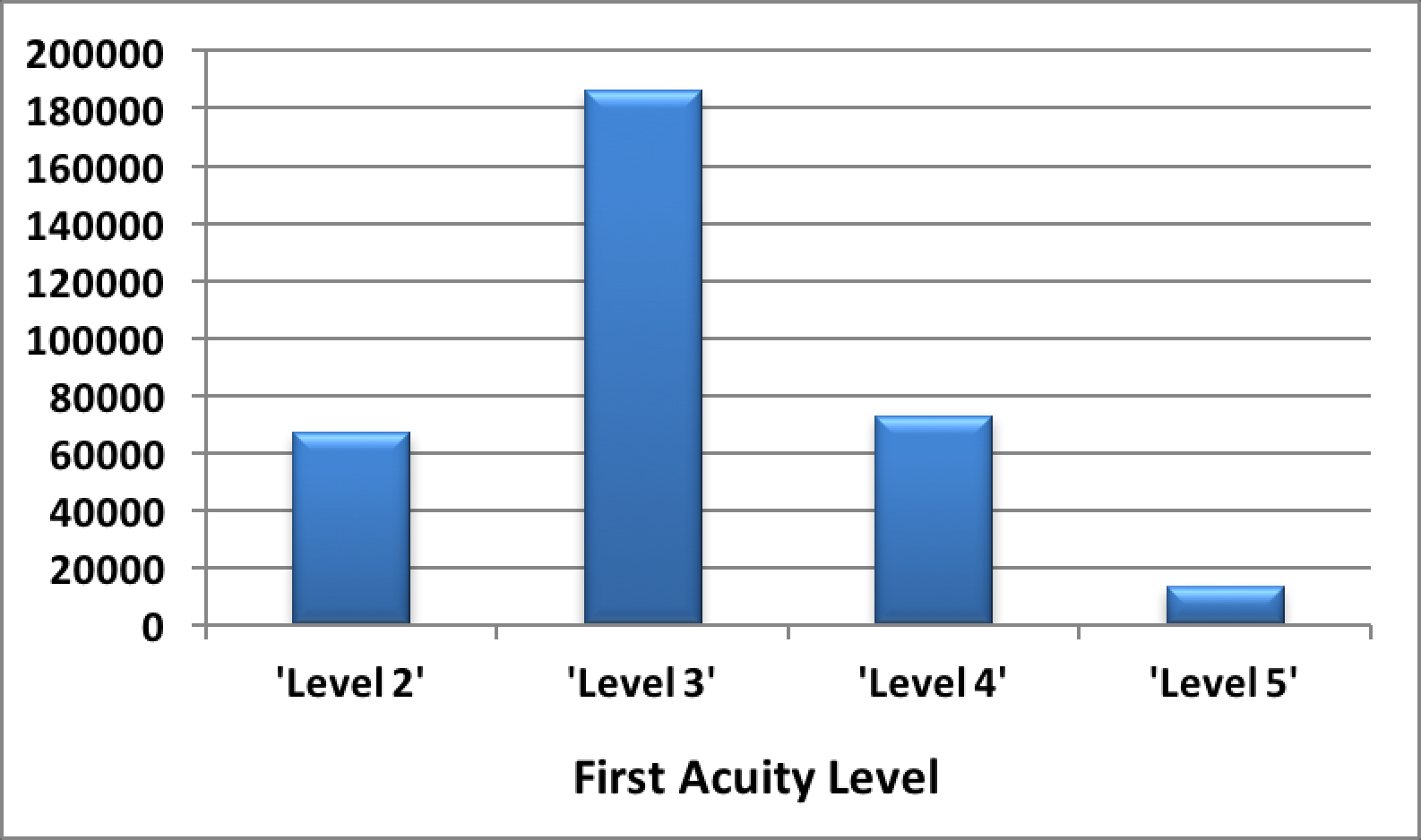}
\label{fig:acuity_level}
}\\
\subfloat[Number of resources for ESI-3 level acuity patients]{
\includegraphics[width=.35\textwidth]{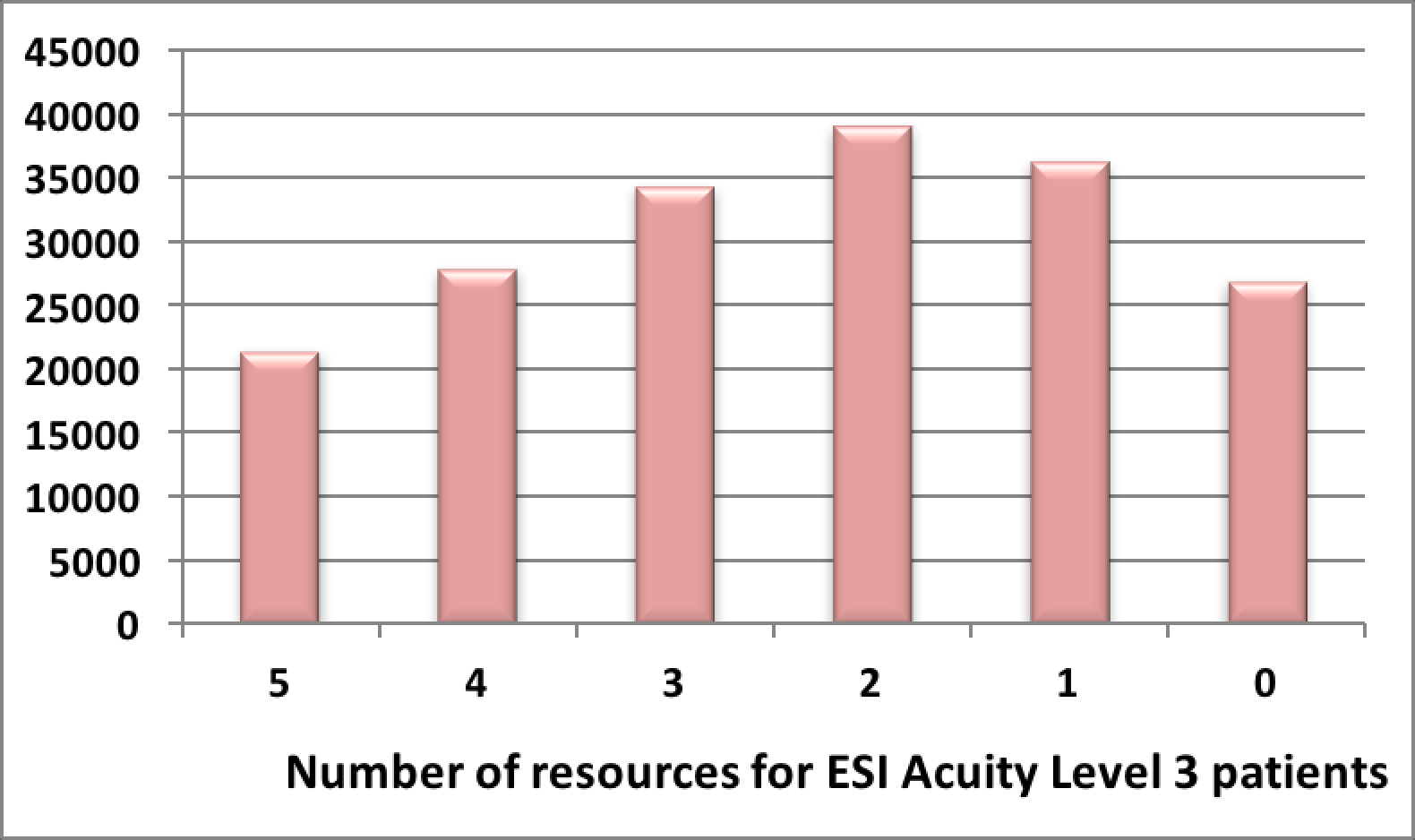}
\label{fig:order_cat_level3}
}
\caption{ESI Level 3 patients prevalence}
\end{figure}

However, the data collected from  338,500 ED visits shows that assessing the number of resources needed for patients' treatments is not that easy task for the nurses.
Namely, Figure~\ref{fig:acuity_level} shows that 55\% of patients were placed into Level 3 ESI category, even though they actually required less than 2 resources in 35\% of the cases (Figure~\ref{fig:order_cat_level3}).
In addition, despite the size and variety of cases belonging to Level 3 ESI category, this group is processed solely in a first-come, first-serve manner.
The order of processing does not take into account within-group resource allocation nor the propensity to have more severe illness, leading to suboptimal organization and utilization of resources. 
\vspace{-5pt}

Therefore, this study focuses on making an efficient and effective system for predicting the number of resources patients would need, in order to provide the ranking among them and assist in patient prioritization and disposition. Specifically, our goal is to utilize routinely available structured data, together with the unstructured textual triage notes and devise a binary-classification model capable of discriminating resource intensive patients from less demanding patients, as well as multi-class classification model for estimating the actual category of number of resources patient would need with better accuracy than triage nurses.

Thus far, natural language understanding presented a huge challenge to convert nurse's notes into structured data useful for prediction of patient visit outcomes \cite{chapman2011,wang2015medical}.
However, given the amount of unstructured textual data that ED information systems contain, utilizing such information
can be a crucial step towards obtaining satisfactory results.
Recent advances in the field of Deep Learning and Natural Language Processing (NLP) have provided powerful approaches for extracting discriminative features from unstructured textual data.
The novel advances proposed different approaches such as neural word embeddings \cite{bengio2003,mikolov2013distributed} or recurrent neural networks (RNNs) \cite{jagannatha2016bidirectional,sutskever2014} that have shown excellent results in many NLP tasks. 

To solve this problem, we propose a novel deep architecture, named Deep Attention Model (DAM). It is built upon bi-directional recurrent neural network (bi-RNN) \cite{schuster1997} suitable for modeling sequence data, such as free text data. 
 Bi-RNN's are not dependent on single direction sequence of information, which is of high importance for texts that are written rapidly and by different persons.
To further mend the issues of triage nurses' noisy text data, we propose the attention based mechanism on top of the bi-RNN word representations to learn attention scores for each word in the notes, which would allow iterpretability of the deep architecture outputs. 
We conduct a thorough experimental evaluation using three years of emergency admission data to examine how the proposed model compare to nurses' performance, as well as current state-of-the art approaches for text classification. We further investigate interpretability of the model by analyzing attention weights on short medical texts.  
Our results suggest potential for substantial improvement of the triage system performance over the current practices.

\section{Related Work}
Firstly, we describe problems and some existing approaches in mining medical text data and then present some notable existing methodological advances used for building highly discriminative features from text which we will exploit for predicting ED admission outcomes.

\subsection{Medical free-text data mining approaches}
Learning from clinical text is a long-standing challenge, and analysis of a such data has become a major focus of research community recently. However, the problem persists as a difficult one, as there are no clear standard methods or tools for analyzing medical text data yet \cite{chapman2011}. In \cite{nguyen2010symbolic} authors predict lung cancer stages from free-text pathology reports. These reports are analyzed using symbolic rule-based approach that used SNOMED\footnote{\url{http://www.snomed.org/snomed-ct}, accessed October 2017} clinical terms to extract key lung cancer characteristics from free-text reports. Approach in \cite{coden2009automatically} predicts cancer and its disease progression from pathology reports using another rule based system for automatic conversion of unstructured pathology reports into a structured data. These models require handcrafting features from text data which is a long and tedious work, and often requires revisiting existing model to improve features. 

To automatically learn features from text data \cite{jagannatha2016bidirectional} proposed to use the bi-RNN algorithm for detection of medical events based on texts from medical records. In their experiments it was shown that bi-RNN significantly outperforms existing tools for medical text analysis.
Approaches for automatic feature extraction are more convenient, do not require handcrafted features and have shown to yield representations of highest quality. Our approach builds on the results of~\cite{jagannatha2016bidirectional} as we use the bi-RNN as one of the building blocks of our model which aims to improve waiting room patient disposition.

\subsection{Neural embedding models for text} \label{sec:embedding_models}
Many advances in modeling sequence data were made in the field of natural language processing, where models for mathematical characterization of language were proposed. Namely, models for distributed low-dimensional representations of words or tokens were initially proposed in \cite{rumelhart1988learning} and were successfully applied more recently in \cite{bengio2003}. Distributed embedding approaches take advantage of word order, and follow the assumption of $n$-gram language models that neighboring words are statistically more dependent. 

Typically, these models learn the probability distribution of the next word given a number of preceding words, which act as the context. 
The probability distribution $\Pr(w_t|w_{t-n+1}:w_{t-1})$ is typically approximated using a neural network \cite{bengio2003} trained to predict a word $w_t$ by projecting the concatenation of vectors for context words $(w_{t-n+1}, \ldots, w_{t-1})$ into a latent representation with multiple non-linear hidden layers and the output softmax layer \cite{bengio2003}. 
More recently, novel architectures (such are continuous bag-of-words (CBOW) and SkipGram that observe both preceding and later words in the sequence) have shown great improvements in representational power and training speed compared to the traditional neural embedding models \cite{mikolov2013distributed}. 

\subsection{Deep Text Models} \label{sec:deep_text_model}
Furthermore, we discuss deeper architectures capable of learning very discriminative and abstract representations of texts.

\subsubsection{Recurrent Neural Networks for text modeling}
The Recurrent Neural Networks (RNNs) models are popular for modeling sequence data. Their power lies in the fact that they maintain an internal state that is updated sequentially which learns representations of word sequences that are used as a proxy for predicting the target, while in previously described approaches, word sequence was often modeled by an order-oblivious sum. Ability to stack multiple layers generates higher order representations that yield great improvements of the model on many tasks. Particular success was achieved using long-short term memory (LSTM) cell as an architecture of RNNs \cite{hochereiter1997}.
More recently, popular sequence-to-sequence paradigm for RNNs was proposed, where input sequence is encoded using the ``encoder'' network, and output sequence is generated using the ``decoder'' network \cite{sutskever2014}. This paradigm has been successfully used for translating sentences from one language into another.

\paragraph{Word Attention Models}
Attention models build upon sequence-to-sequence (encoder-decoder) paradigm of RNNs by dynamically re-weighting (i.e. focusing attention) on various elements of the source (text) representation during the decoding process, and they have demonstrated considerable improvements over their non-attention counterparts \cite{bahdanau2014}. Attention mechanism was developed as a separate neural network that takes sequence of word embeddings and learns attention scores for each word, with higher attention assigning to more ``important'' words in the document leading to more focused higher order representation of the sequence. Attention models have more recently been adapted for the general setting of learning compact representations of documents \cite{zhai2016deepintent}. 

\paragraph{bi-LSTM}
Another interesting paradigm are bi-directional recurrent neural networks, where two RNNs (i.e. LSTM, thus bi-LSTM), encode the text as a forward and backward sequence, respectively \cite{schuster1997}. Final words representation is obtained by concatenating representations of the two LSTMs and it was observed that bi-LSTM's perform well on datasets without the strict order in sequences, such is the case with triage text data. 

\subsubsection{Convolutional Neural Networks for text}
\label{sec:cnn_for_text}
Recently, architectures for sequence modeling increasingly include temporal convolutions as building blocks. A good example of such approach is ConvNet for text classification \cite{zhang2015character} and Very Deep CNN (VDCNN) model \cite{conneau2016very}, both of which are using temporal convolutions to model sequence of words (or characters) with task to perform classification. These models have been successful to outperform RNN based models. In this study, we will use word-level VDCNN as our primary baseline. 

It should be noted that the above mentioned convolutional approaches are designed for character level modeling, while we use them for word level modeling. Our reasons are that medical notes do not generate sufficient amount of data for character level models to learn proper mappings, even though advantages of such approaches can be very useful in our setting: Nurses use many abbreviations, make many typos during triage process, etc. Our initial experiments on character level modeling were unsatisfactory, and further pursue for such approach will be left for future work.

\section{Proposed Model}
\label{sec:proposed_model}
\begin{figure}[h!]
	\centering
	\captionsetup{justification=centering}
	\includegraphics[width=0.9\linewidth]{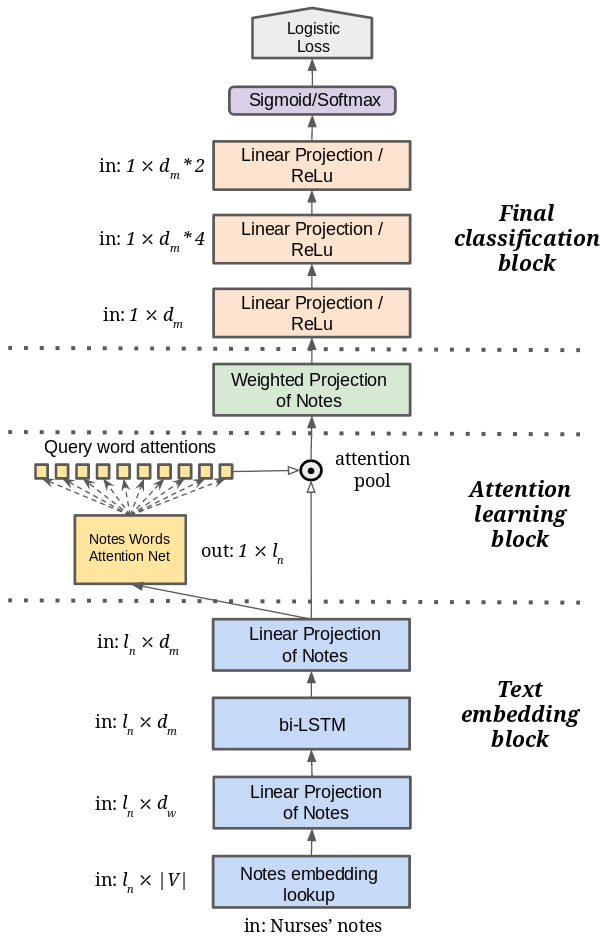}
	\caption{Deep Attention Model Architecture}
	\label{fig:dam_model}
\end{figure}

The architecture of the triage acuity prediction model, which we refer to as Deep Attention Model (DAM), is presented in Figure~\ref{fig:dam_model}. Before exploring its architecture, we first explain how the input is represented.
\paragraph{Text embedding block}
Let us assume that we are given a set $\mathcal{P}$ of patient records, where each patient record $p_n$ contains an unstructured portion of data $p_n^u$, and a structured portion of data $p_n^s$ and an outcome category $c_n$. Unstructured portion of the data contains text on chief complaint, medical history, home medications and nursing notes, while structured portion of the data includes routinely available continuous and nominal data such as hearth rate, blood pressure, temperature, patient age group and other.

In order to extract meaningful features from unstructured portion of the data, we consider $p_n^u$ of length $l_n$ as a document containing a sequence of words $(w_1, w_2, \ldots, w_{l_n})$. Document length $l_n$ is maintained via null character padding and cropping. We first embed each word in the document $p_n^u$ into a $d_w$ dimensional vector using word embedding layer. To learn different interaction between words as they might repeat or correlate in the sections of the triage nurses' notes we pass our embeddings through a dense layer to obtain $d_m$ dimensional word vectors. These cross learned word vectors are then passed through bi-LSTM layer that learns sequence dependency of words in both directions. bi-LSTM layer has the capability to mend the bias in taking the notes, as some nurses may ask for chief complaint first, some might start from previous medical history and some might ask when did the symptoms occur first before asking for chief complaint if symptoms are visually identifiable, etc. Learning sequence from both directions can capture relations between keywords across text.
Finally, one more fully connected layer is used to embed words to capture higher order nurses' notes embeddings. It should be noted that without using this dense layer and using only bi-RNN embeddings, model persistently yielded poorer results. Last two layers in text embedding block have an unchanged embedding size $d_m$ in our experiments. 

\paragraph{Attention block}
The output of previous layers is a matrix $h_n$ of size $l_n \times d_m$, where we embed every word in a $d_m$ dimensional space. It is desirable to compress this matrix representation into a single vector to make it easy to build a loss function and facilitate training. 

This can be achieved in multiple ways, which we refer to as pooling, i.e. one can take the sum of all vectors in the document ($\sum_{i}h_n^{(i)}$) for sum pooling, take average ($\frac{1}{l_n}\sum_{i}h_n^{(i)}$) for average pooling or take maximum ($\max(h_n^{(1)},\ldots,h_n^{(l_n)})$) for max pooling. However, these pooling strategies do not take into account ``importance'' of different words, where it is desired to give more weight to words that are more important (or provide more information) in the text. In our experiments sum-pooling was always the best performing strategy and we will use it as a baseline.

To evaluate the importance of different words in document, we adopt the techniques from machine translation, namely sequence to sequence learning \cite{bahdanau2014} and adapt them to a more general case where compact representations are needed \cite{zhai2016deepintent}. 
The advantage of attention models, is that word attention scores $a^{(i)}$ are generated dynamically based on the given context, and as such are independently obtained for each document. These scores are obtained using a separate neural network architecture $s(h_n^{(i)}; \theta)$ that simply learns to score words based on their embeddings using softmax function:
{\small\begin{equation}
a{(i)} = \frac{exp(s(h_n^{(i)}; \theta))}{\sum_{i = 1}^{l_n} exp(s(h_n^{(i)}; \theta))}.
\end{equation}}
Neural network $s(h_n^{(i)}; \theta)$ (two fully connected layers with ReLU nonlinearities in our experiments) learns real valued scores, normalized across the document, given document representation $h_n$. Learning embeddings is coupled with the entire network in our model, allowing for an end-to-end training. 
The final vector projection of nurses' notes is then obtained as $v_n = \sum_{i}a^{(i)}*h_n^{(i)}$.

Finally, learned attentions allow the model to focus on more important words, as we will evaluate in Section~\ref{sec:attention_analysis}. This enables interpretability of the scores provided by the model, which is mandatory for decision making in the emergency departments.

\paragraph{Final block}
Obtained vector representation of the notes $v_n$ is then fed into three fully connected layers with ReLU activations as shown in Figure~\ref{fig:dam_model}. Final layer is used for scoring of classes and sigmoid function is used for binary classification and softmax for multiclass classification, which provide scores $p_n$.

We choose logistic loss function to optimize the model:
{\small 
\begin{equation}
\mathcal{L} = \sum_{i \in |C|}(\sum_{n: c_n^i = 1}\log p_n^i + \sum_{n: c_n^i = 0} \log (1- p_n^i)),
\end{equation}}
where $|C|$ is the set of possible classes that depends on whether we want to split patients into two cohorts of higher and lower acuity patients (binary classification) or want to predict actual resource category (multiclass classification). 

\paragraph{Adding handcrafted features}
As mentioned before, triage systems collect vast amount of structured data $p_n^s$ as well. To incorporate such data into our model, we append vector of features collected for the patient to the final layer of embeddings. This creates a deep and wide \cite{cheng2016widedeep} architecture of the DAM model.

\section{Experiments}
\label{sec:experiments}
Here we show evaluations of the proposed approach. We first present how we construct the data set to conduct the experiments, describe different baselines that are relevant for this study, and provide parameters used in our experiments. We then dive into research questions raised in the introduction and discuss the results obtained from the related experiments. 
\subsection{Experimental set-up}
\subsubsection{Triage Data}
\label{sec:data}
The data used in this study is retrospective review of ED data over a 3 year period 2012.--2015. at an urban academic medical center. 
Routinely available continuous and nominal data includes: ED assigned location, gender, age range, method of arrival, hour of arrival, number of prior ED visits, insurance group, heart rate, systolic blood pressure, and temperature. Data are binarized to create binary vector $p_s$. Text data included chief complaint, past medical history, medication list, and free text initial nursing assessment.
The response variable represents category of number of resources (0-5) for multi-class classification task, and it is binarized as positive class for the patients who consumed 3, 4 or 5 resources vs. those who consumed 0 through 2 resource categories.
The training population in our experiments consists $250,000$ ED visits, while  $20,500$ patient cases were extracted as a validation set.
The test set is comprised of $68,500$ patients out of which $36,883$ are ESI level 3.


\subsubsection{Baselines}
In the experiments we evaluate quality of predictions when using different data sources. 
Thus, our first choices of baselines are the logistic regression (LogReg) and the multi-layer perceptron (MLP) models that use only handcrafted structured features.

In order to learn features from text data, we employ a basic word embedding model (embd) where we learn word embeddings through classification framework in an end-to-end manner.
As suggested in the literature \cite{jagannatha2016bidirectional} we compare to bi-LSTM model which was successfully applied for analysis of medical texts in the past.
For a representative of very deep text models we employ the word-level VDCNN model described in Section~\ref{sec:cnn_for_text}.
Finally, to evaluate improvements of attention layer we employ sum pooling on our model (annotated as DSMP) as discussed in Section~\ref{sec:proposed_model}.
All deep models are evaluated with and without using handcrafted features to investigate whether using them provides lift in accuracy.

\subsubsection{Experimental platform}
We implemented all the algorithms using Tensorflow platform and run them on a machine with two Nvidia Tesla P100 GPUs with 16 GB RAM and Intel Xeon CPU with 512GB RAM. 
Adam optimizer is used to minimize loss, with 0.001 starting learning rate, and batch size of 512 examples. 
We use a held out validation set to monitor the training progress, and all the models are trained till the validation loss stops decreasing.
Dimensionality of notes word embeddings is set to 300, while for bi-LSTM, first and final linear projection embeddings is set to 200. For VDCNN model, we used 64 filters in convolutional layers.

\subsection{Experimental Results}
In this section we report the performance of the proposed and baseline approaches for the number of resources category prediction tasks as binary and multi-class classification problems in terms of ROC AUC and accuracy for all the models. 
We aim to answer the following research questions:

\subsubsection{Research Question 1:}
\textit{Can we automatically learn representations from the nurses' notes textual content, without any feature handcrafting, in order to predict the number of resources category? Are additional structured features helpful for this task?
}

Our objective is to test to what extent the text data representations learned by the introduced deep models, are effective for the number of resources category prediction task and how they compare to the models that use only structured data. In addition, we examine how they perform when using both structured and text data. 
\begin{table}[]
\small{
\centering
\caption{Comparison of models that utilize only structured data, against DAM models trained on only unstructured, and on both types of input data.}
\label{tbl:deep_ws_wide}
\begin{tabular}{l|c|c|c|c|}
\cline{2-5}
\multicolumn{1}{l|}{}                     & \multicolumn{2}{c|}{\textbf{Binary}} & \multicolumn{2}{c|}{\textbf{Multi-class}} \\ \hline
\multicolumn{1}{|c|}{\textbf{Models}}     & \textbf{Acc.} & \textbf{AUC} & \textbf{Acc.}  & \textbf{AUC} \\ \hline
\multicolumn{1}{|c|}{\textbf{LogReg}}     & 54.91\%           & 0.5277           & 16.34\%            & 0.4982               \\ \hline
\multicolumn{1}{|c|}{\textbf{MLP}}        & 56.13\%           & 0.5689           & 19.88\%            & 0.5027               \\ \hline
\multicolumn{1}{|c|}{\textbf{DAM-$p_u$}}   & \textbf{79.25\%}  & 0.8763           & 43.30\%            & 0.6680               \\ \hline
\multicolumn{1}{|c|}{\textbf{DAM-$p_u$,$p_s$}} & 79.21\%           & \textbf{0.8797}  & \textbf{43.80\%}   & \textbf{0.6715}      \\ \hline
\end{tabular}
}
\end{table}
We compare learning models that can be implemented using the existing structured data, and models capable of utilizing additional unstructured text data.
Table~\ref{tbl:deep_ws_wide} shows that the DAM outperforms LogReg (AUC 53\% for binary and 50\% for multi--class), as well as MLP model (AUC 57\% for binary and 50\% for multi-class) learned on structured data only. 
From the results, we see that DAM was able to learn representation from the triage notes textual content, and predict the number of resources category much better (AUC 88\%  for binary and 67\% for multi-class classification task) than the models that learned from structured data only (31\% and 17\% lift in AUC, respectively).
Even though representations learned from text bring majority of predictive performance improvement, we note, however, that DAM model learned on both text data and structured data performed slightly better than DAM model learned on text data only, showing that both of the data sources are informative and should be used in decision making. 


\subsubsection{Research Question 2:}
\textit{How do the DAM models perform for the task of identifying more resource-intensive patients vs. less?
How do they compare to the baseline models on this binary classification task (more than 2 resources need as positive class or 2 or less resources needed as negative class)?
}

To answer this question we compare our proposed DAM model with baselines able to learn from both structured and unstructured data simultaneously. Table~\ref{tbl:binary_claass} presents the predictive performance on binary classification task, and it can be seen that our proposed DAM outperformed the alternatives. Slightly less accurate is the DSMP model, which is the version of DAM where the attention block is replaced with the sum pooling strategy. Even though attention block appears to bring a small improvement in generalization ability, its main benefit is that it introduces interpretability, which we will investigate in research question 5. Primary baselines VDCNN and bi-LSTM models are few percent less accurate than the DAM, while linear embedding approach performs considerably worse.

\begin{table}[]
\centering
\small{
\caption{DAM vs baselines performances for the binary classification task (more than 2 resources, or less).}
\label{tbl:binary_claass}
\begin{tabular}{c|c|c|c|c|}
\cline{2-5}
                                      & \multicolumn{2}{c|}{\textbf{Accuracy}} & \multicolumn{2}{c|}{\textbf{ROC AUC}} \\ \hline
\multicolumn{1}{|c|}{\textbf{Models}} & \textbf{$p_u$}        & \textbf{($p_u$,$p_s$)}    & \textbf{$p_u$}       & \textbf{($p_u$,$p_s$)}    \\ \hline
\multicolumn{1}{|c|}{\textbf{embd}}   & 55.30\%            & 64.33\%           & 0.5165            & 0.6155            \\ \hline
\multicolumn{1}{|c|}{\textbf{bi-LSTM}} & 76.59\%            & 77.08\%           & 0.84626           & 0.8523            \\ \hline
\multicolumn{1}{|c|}{\textbf{VDCNN}}  & 76.81\%            & 77.70\%           & 0.8467            & 0.8609            \\ \hline
\multicolumn{1}{|c|}{\textbf{DSMP}}   & \textit{78.79\%}   & \textit{78.63\%}  & \textit{0.8713}   & \textit{0.8717}   \\ \hline
\multicolumn{1}{|c|}{\textbf{DAM}}    & \textbf{79.25\%}   & \textbf{79.21\%}  & \textbf{0.8763}   & \textbf{0.8797}   \\ \hline
\end{tabular}
}
\end{table}

\subsubsection{Research Question 3:}
\textit{How do the DAM models perform for the task of exact number of resources category prediction? How do they compare to the baseline models on this multi-class classification task (classes are number of resources category - 0,1,2,3,4,5)?
}

We evaluate all models on the task of multiclass classification, where classes are number of resources used by the patient (where 5 or more resources are assigned to the same category). 
Results are provided in Table~\ref{tbl:multi_claass}, with accuracy and average AUC (averaged over one-vs-all evaluation approach). As in the binary case, we observe that DAM is the best performing model with second best model is sum pooling alternative to attention layer. bi-LSTM and VDCNN approaches are best performing baselines, VDCNN being slightly better, and are consistently outperformed by the DAM. Furthermore, we can see that attention persistently provides improvement over next best pooling strategy, yielding best performing model that is also interpretable. 
As in the binary case, we can see that using structured data stably helps in improving prediction quality.

We next examine how does the DAM model compare to human performance on multi-class classification, as this is the exact task triage nurses are ask to perform (in the step III).


\begin{table}[]
\centering
\small{
\caption{DAM vs baselines performances for the multi-class classification task (number of resource category).}
\label{tbl:multi_claass}
\begin{tabular}{cc|c|c|c|}
\cline{2-5}
\multicolumn{1}{l|}{}                 & \multicolumn{2}{c|}{\textbf{Accuracy}} & \multicolumn{2}{c|}{\textbf{Average AUC}} \\ \hline
\multicolumn{1}{|c|}{\textbf{Models}} & \textbf{$p_u$}           & \textbf{($p_u$,$p_s$)}    & \textbf{$p_u$}           & \textbf{($p_u$,$p_s$)}        \\ \hline
\multicolumn{1}{|c|}{\textbf{Embd}}   & 14.74\%               & 14.74\%           & 0.5000                & 0.5000                \\ \hline
\multicolumn{1}{|c|}{\textbf{bi-LSTM}}  & 39.16\%               & 38.50\%           & 0.6401                & 0.6358                \\ \hline
\multicolumn{1}{|c|}{\textbf{VDCNN}}  & \textit{39.68\%}      & \textit{41.33\%}  & 0.6390                & \textit{0.6506}       \\ \hline
\multicolumn{1}{|c|}{\textbf{DSMP}}   & 39.61\%               & 40.67\%           & \textit{0.6412}       & 0.6494                \\ \hline
\multicolumn{1}{|c|}{\textbf{DAM}}    & \textbf{43.30\%}      & \textbf{43.80\%}  & \textbf{0.6680}       & \textbf{0.6715}       \\ \hline
\end{tabular}
 }
\end{table}

\subsubsection{Research Question 4:}
\textit{How does the DAM model performance compare to nurses' performance in the number of resources category prediction task?}

As ESI levels represent a surrogate for actual number of resources category, from the available data we can only approximate nurses' performance on this task. 
We approximate nurses' prediction accuracy by analyzing how many patients fall in a particular (\textit{First Acuity Level}, \textit{Number of Resources Category}) group. This is a reasonable approach to evaluate their performance because of the way ESI levels are assigned (Figure~\ref{fig:ESI}). Namely, in the step III nurses should predict how many resources a patient would require,
and
assign that patient to ESI Level 5 if it doesn't require any resources, 
Level 4 if they require 1 resource, 
Level 3 if they require more resources, 
or consider Level 2 if the vitals are in danger zone (and they might require even more resources).
Having this procedure in mind, we group number of resources categories to match ESI acuity levels as in Table~\ref{tbl:ESI_ORDCAT}.
\begin{table}[h!]
	\centering
    \footnotesize{
	\captionsetup{justification=centering}
	\caption{Approximating number of resources category by first acuity level prediction}
    \label{tbl:ESI_ORDCAT}
	 \begin{tabular}{c|c}
		Number of Resources Category  & First Acuity Level  \\
		\hline
        0 & Level 5 \\
        1 & Level 4 \\
        2 or 3 & Level 3 \\
        4 or 5 & Level 2 \\
	\end{tabular}
    }
\end{table}

Then we approximate the nurse performance for the task of number of resources category prediction via their first acuity level assessment. Table~\ref{fig:nurses_performance} presents these measurements on the whole dataset. We observe that majority (55\%) of the patients were predicted to fall within Acuity Level 3 category, even though they needed only 1 (10\%) or even 0 (8\%) resources, and that only 10\% of patients that did not need any ED resources (recall=0.10) were assigned Level 5 category. 
On the other hand, patients that required 4 or more resources (14.5\%) were possibly misplaced within lower category as they required more resources than most of the patients within Level 3 category and should have higher priority.
All of the above leads to total approximated accuracy of \textbf{43.6\%}.
\begin{table}[h!]
  \centering
	\captionsetup{justification=centering}
	\includegraphics[width=\linewidth]{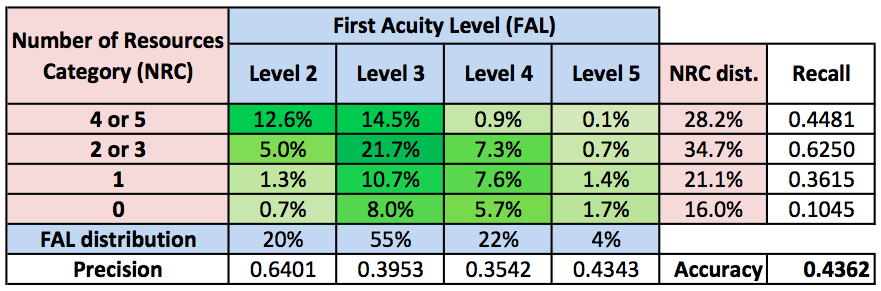}
	\caption{Nurses' performance approximation for the task of \textit{Number of Resources Category} prediction via their \textit{First Acuity Level} assessment} 
	\label{fig:nurses_performance}
\end{table}
\vspace{-10pt}
By using the DAM algorithm, we were able to reduce off-diagonal patient counts (Table~\ref{fig:dam_conf_matrix}). The distribution of predicted resources groups better matches true distribution of the test data (ie. 15\% to 16\% for 0 resources, 20\% to 21\% for 1 resource, 36\% to 32\% for 2 or 3 resources and 29\% to 31\% for 4 or more resources) giving us the total accuracy of $\sim$~\textbf{\ 60\%} and a potential lift of 16\% in accuracy compared to nurses' assessments. 
\begin{table}[h!]
	\centering
	\captionsetup{justification=centering}
	\includegraphics[width=\linewidth]{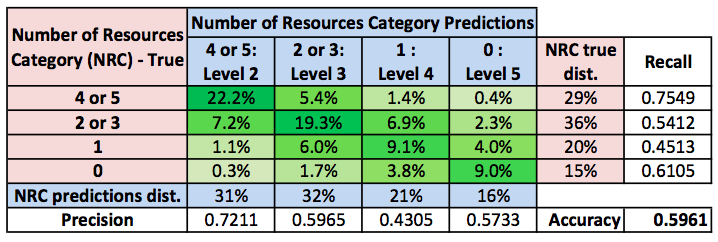}
	\caption{DAM performance for the task of \textit{Number of Resources Category} prediction (grouped by the First Acuity Level approximation groups)} 
	\label{fig:dam_conf_matrix}
\end{table}
\vspace{-10pt}
\subsubsection{Research Question 5:}
\label{sec:attention_analysis}
\textit{Can the proposed DAM model be leveraged to improve the existing triage system flows? How can the attention weights be utilized by medical practitioners to understand predictions?}

Finally, we provide comments on how a system like this can be implemented in actual hospital emergency department systems. The ability of the DAM to discriminate high from lower priority patients as well as to accurately predict number of resources a patient will need allows it to be used for improving patient disposition in the waiting rooms. Such tool can aid triage nurses and doctors to prioritize patients accurately and with minimal bias, especially in uncertain cases, providing the service to the patients that need it the most.

\begin{figure}[h]
\centering
\captionsetup{justification=centering}
\includegraphics[width=\columnwidth]{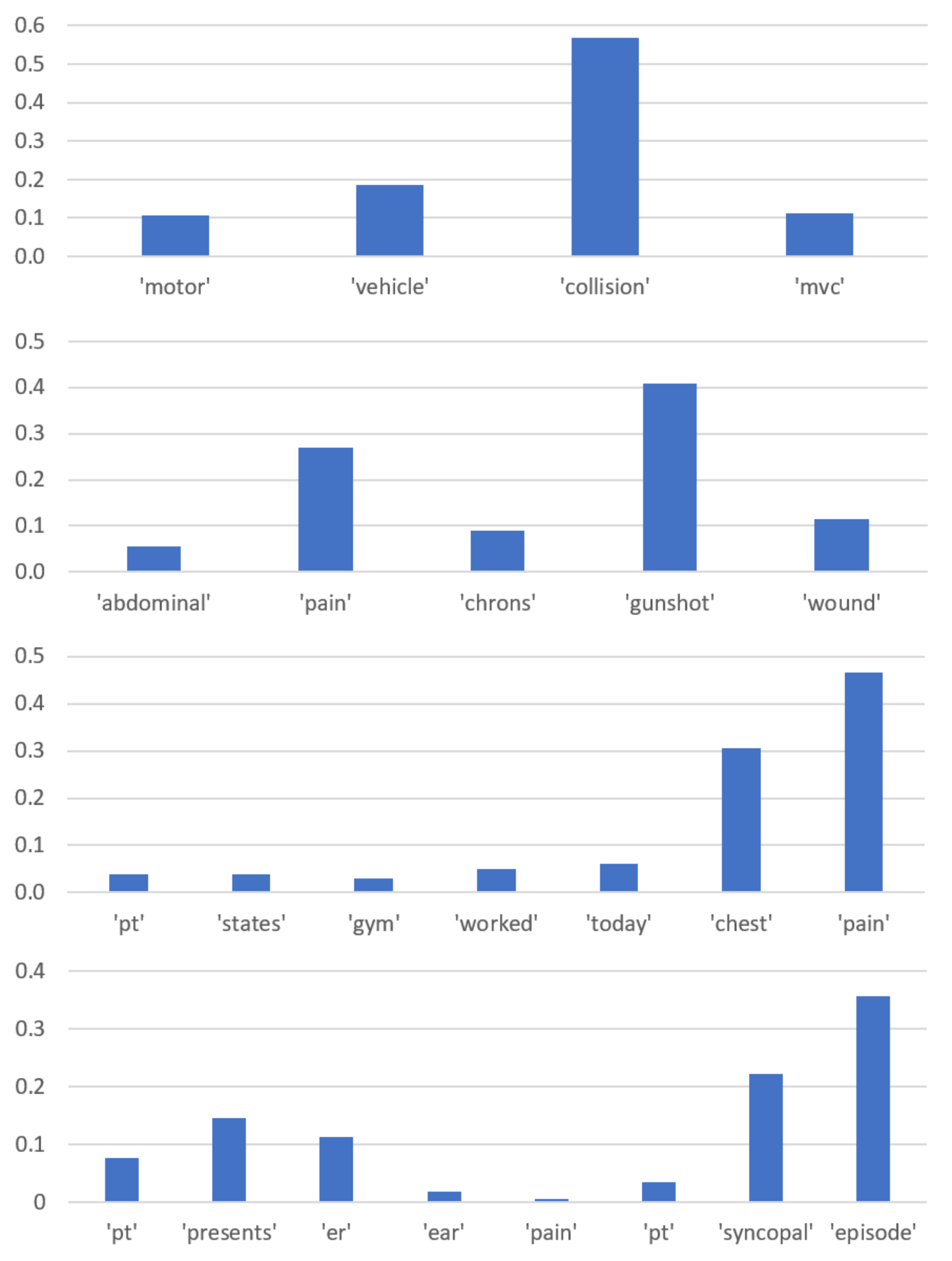}
\caption{Examples of patient complaints during triage process with attentions.}
\label{fig:attentions}
\end{figure}
\vspace{-5pt}
In order for computer assisted triage to be accepted by experienced triage staff, having simple estimation without useful clinical feedback will make its deployment more difficult. Thus, being able to provide some intuition on how was the estimation obtained is mandatory. In the DAM, attentions learned for words in notes can act as a proxy for such intuition. Higher weights can tell triage staff what was factored in the given prediction, thus allowing control over potential errors model can make. Ultimately, if everything is rendered satisfying by the staff, actionable decision can then be taken. 
To evaluate attentions of the DAM, we generate several examples of patients complaint and run trained algorithm to evaluate the attentions (Figure~\ref{fig:attentions}). We can see that model is capable of focusing on the key words of patients complaint such as: ``vehicle ``collision'', ``pain'' and ``gunshot'', ``chest pain'', ``syncopal'', etc., while assigning less of attention to other keywords that might be repeating: ``mvc'', which stands for motor vehicle, or non-critical like ``pt'', which is abbreviation for patient.
It is difficult to properly quantify the quality of obtained attentions, however in the deployed system, triage staff can be allowed to grade attentions for each case thus allowing for supervision in retraining model to obtain higher quality attention mechanism. This will be pursued in the future work.

\vspace{-5pt}
\section{\textsc{Conclusion}}
\label{sec:conclusion}
In this study we addressed the problem of high variance subjective resource utilization outcomes prediction in triage rooms. For this task we show that utilizing nurses' notes can provide a significant improvement in accuracy compared to standard continual and nominal data. We proposed a novel model to exploit medical texts and obtain state-of-the-art predictive accuracy, finally outperforming reported accuracies of triage staff. Attentions the proposed model learns can be very useful in providing clear feedback on what guided the predictions aiding interpretability and clinical acceptance of the model. We further aim to address several more issues that nurses' notes data have, such are common typos, big variety of abbreviations and human bias.
\vspace{-5pt}
\section{\textsc{Acknowledgments}}
The authors would like to thank Nemanja Djuric and Vladan Radosavljevic for their valuable comments. This research was supported in part by the NSF through major research instrumentation grant number 1625061 and Pennsylvania Department of Health CURE Health Data Science Research Project.
\vspace{-10pt}

\end{document}